\documentclass[submission, Proceedings]{SciPost}

\hypersetup{
    colorlinks,
    linkcolor={red!50!black},
    citecolor={blue!50!black},
    urlcolor={blue!80!black}
}

\usepackage[bitstream-charter]{mathdesign}
\def\ie{{\it i.e.}}

\newcommand{\NLOstar}{NLO$^\star\,$}
\def\jpsi    {\mbox{$J/\psi$}}

\DeclareSymbolFont{usualmathcal}{OMS}{cmsy}{m}{n}
\DeclareSymbolFontAlphabet{\mathcal}{usualmathcal}

\newcommand{\Q}{\mathcal{Q}}
\usepackage{amsmath}
\usepackage{enumerate}
\usepackage{hyperref}

 \hypersetup{
   pdftitle={},%
   pdfauthor={},%
   pdfsubject={},%
   pdfkeywords={},%
   pdfstartview={},%
   bookmarksopen=true, breaklinks=true, debug=true, %
   colorlinks=true, linkcolor=red, citecolor=blue, urlcolor=blue
 }

\begin{document}

\begin{center}{\Large \textbf{
NLO inclusive $J/\psi$ photoproduction at large $P_T$ at HERA and the EIC
}}\end{center}

\begin{center}
C.~Flore\textsuperscript{1$\star$},
J.-P.~Lansberg\textsuperscript{1},
H.-S.~Shao\textsuperscript{2} and
Y.~Yedelkina\textsuperscript{1}
\end{center}

\begin{center}
{\bf 1} Universit\'e Paris-Saclay, CNRS, IJCLab, 91405 Orsay, France
\\
{\bf 2} Laboratoire de Physique Th\'eorique et Hautes Energies, UMR 7589,
Sorbonne Universit\'e et CNRS, 4 place Jussieu, 75252 Paris Cedex 05, France
\\
* carlo.flore@ijclab.in2p3.fr
\end{center}

\begin{center}
\today
\end{center}


\definecolor{palegray}{gray}{0.95}
\begin{center}
\colorbox{palegray}{
  \begin{tabular}{rr}
  \begin{minipage}{0.1\textwidth}
    \includegraphics[width=22mm]{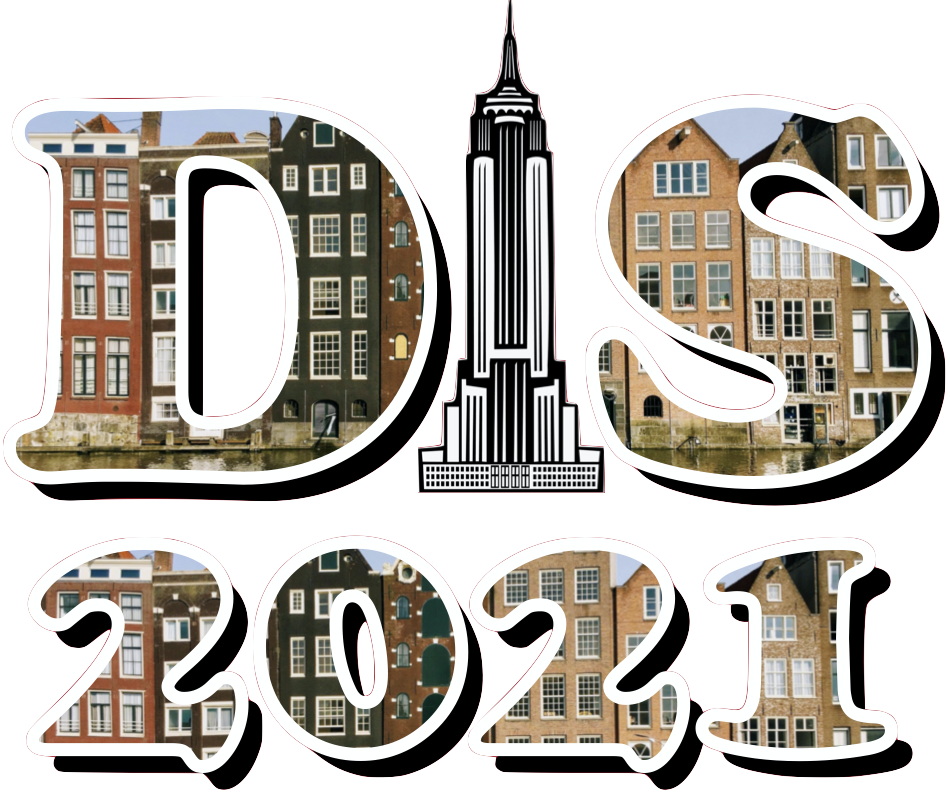}
  \end{minipage}
  &
  \begin{minipage}{0.75\textwidth}
    \begin{center}
    {\it Proceedings for the XXVIII International Workshop\\ on Deep-Inelastic Scattering and
Related Subjects,}\\
    {\it Stony Brook University, New York, USA, 12-16 April 2021} \\
    \doi{10.21468/SciPostPhysProc.?}\\
    \end{center}
  \end{minipage}
\end{tabular}
}
\end{center}

\section*{Abstract}
{\bf
We study inclusive $J/\psi$ photoproduction at NLO at large $P_T$ at HERA and the EIC. Our computation includes NLO QCD leading-$P_T$ corrections, QED contributions via an off-shell photon as well as those from $J/\psi$+charm channels. For the latter, we employ the variable-flavour-number scheme. Our results are found to agree with the latest HERA data by H1 and provide, for the first time, a reliable estimate of the EIC reach for such a measurement. Finally, we demonstrate the observability of $J/\psi$+charm production and the sensitivy to probe the non-perturbative charm content of the proton at high $x$, also known as intrinsic charm, at the EIC.
}

\section{Introduction}
\label{sec:intro}
Inclusive $\jpsi$ photoproduction, when an almost on-shell photon hits and breaks a proton producing the $\jpsi$, is a useful tool for understanding the quarkonium-production mechanism and to learn more about the gluon content of the proton. 

In Ref.~\cite{Flore:2020jau}, we presented a comprehensive analysis on the NLO $\jpsi$ photoproduction in the large-$P_T$ region ($P_T \gg M_{J/\psi }$) which was seldom studied at HERA~\cite{Aid:1996dn,Breitweg:1997we,Chekanov:2002at,Adloff:2002ex,Chekanov:2009ad,Aaron:2010gz,Abramowicz:2012dh}, and which could be experimentally studied in more details at the future US Electron-Ion Collider (EIC)~\cite{Accardi:2012qut}. The analysis is based on the Colour-Singlet Model (CSM)~\cite{Berger:1980ni}, which is the leading-$v$ contribution of Non-Relativistic QCD (NRQCD)~\cite{Bodwin:1994jh}. 

At variance with previous works on inclusive $\jpsi$ photoproduction, where only full QCD contributions were considered, we also included two, so far overlooked, contributions: the associated $\jpsi$+charm production, and the pure QED contribution, where the $\jpsi$ is produced by an off-shell photon, that happens to be relevant at large $P_T$. The full list of the considered partonic subprocesses is the following:
\begin{enumerate}[(a)]\setlength{\itemsep}{.1mm}
 \item $\gamma + g \to \jpsi + g (+ g)$;
 \item $\gamma + \{q,\bar{q}\} \to \jpsi + \{q,\bar{q}\} (+ g)\,$ ($q = u, d, s$);
 \item $\gamma + g \to \jpsi + c  + \bar{c}$ and $\gamma + \{c,\bar{c}\} \to \jpsi + \{c,\bar{c}\}$.
\end{enumerate}
$\gamma + g \to \jpsi + g$ represents the $\mathcal{O}(\alpha \alpha_s^2)$ LO QCD contribution, while $\gamma + q \to \jpsi + q$ is the $\mathcal{O}(\alpha^3) $ LO QED subprocess. The $\mathcal{O}(\alpha \alpha_s^3)$ contributions are given by $\gamma \{g,q,\bar{q}\} \to \jpsi + \{g, q,\bar{q}\} $ $+ g$. The sum of $\mathcal{O}(\alpha \alpha_s^2)$ and $\mathcal{O}(\alpha \alpha_s^3)$ terms is the NLO QCD contribution. Finally, we note that the partonic subprocesses (c) can be calculated respectively in the 3  and 4 Flavour Schemes (3FS and 4FS). A proper treatment of such channels is required, and is given by the LO Variable Flavour Number Scheme (LO VFNS) illustrated in Section 2.4 of Ref.~\cite{Flore:2020jau}, to which we guide the reader for more details.

\section{$\jpsi$ photoproduction at finite $P_T$ and the \NLOstar approximation}

We first start by recalling some elements of kinematics. We define $s_{ep}=(P_e + P_p)^2=4 E_e E_p$ ($E_{e(p)}$ is the electron (proton) beam energy) and $s_{\gamma p}=W_{\gamma p}^2 = (P_\gamma + P_p)^2$. Introducing $x_\gamma$ as $P_\gamma= x_\gamma P_e$, it turns out that $s_{\gamma p}=x_\gamma s_{ep}$. Another important variable, called elasticity, is defined as $z = (P_{\Q}\cdot P_p)/(P_\gamma \cdot P_p)$. $z$ reduces to the photon energy taken by the $\jpsi$ in the proton rest frame, and can also be expressed as $z=( 2\,E_p \,m_T)/(W^2_{\gamma p}\,e^{y})$ in terms of the $J/\psi$ rapidity $y$ (with $y$ and $E_p$ defined in the same frame) and  the quarkonium transverse mass, $m_T = \sqrt{m_\Q^2 + P_{\Q T}^2}$.  

In our evaluation of the NLO corrections to $J/\psi+g$, we resorted to the NLO$^\star$ approximation~\cite{Artoisenet:2008fc,Lansberg:2008gk}, which allows one to consider the leading-$P_T$ contributions by imposing a democratic,  lower cut on the invariant mass of every pair of massless partons, denoted by $s_{ij}^{\rm min}$. Such an approximation has already been tested in the case of hadroproduction (see \textit{e.g.}~Ref.~\cite{Lansberg:2008gk}), and it is automated within {\sc HELAC-Onia}~\cite{Shao:2012iz, Shao:2015vga}, that we used for our computations. As for any leading-$P_T$ topology associated to real NLO emission $s_{ij}$ for any $i,j$ pair grows with $P_T$, the result becomes insensitive to $s_{ij}^{\rm min}$, and only the subleading $P_T$ contributions exhibit a $\log(s_{ij}^{\rm min})$ dependence which is thus $P_T$-power suppressed. 

For testing the validity of the NLO$^\star$ approximation, we checked our approximate calculation against a full NLO calculation by Butensch\"on and Kniehl~\cite{Butenschoen:2011yh}. It turned out that a suitable interval for our analysis is $\sqrt{s_{ij}^{\rm min}}/m_c \in \left[1:3\right]$ ($m_c = 1.5$ GeV), and that $\sqrt{s_{ij}^{\rm min}} = 2 m_c$ remarkably reproduces the complete NLO calculation. The results are shown in Fig.~2 of Ref.~\cite{Flore:2020jau}. In view of such a result, we revisited the latest HERA data from H1 Collaboration and then extended our computation to the case of the future EIC.

\section{Results}

In the following, we present our results for HERA and the future EIC. In every plot, we show the following CS contributions: LO (blue) and \NLOstar QCD (red), LO QED (light blue), LO VFNS (green) and the overall sum (orange). A feed-down contribution $\psi^\prime \to \jpsi$ of $20\%$ is also taken into account\footnote{We guide the reader to Section 3 of Ref.~\cite{Flore:2020jau} for a detailed discussion about the considered feed-downs.}.

For our predictions, we used the CT14NLO proton PDFs set~\cite{Dulat:2015mca}. The corresponding theoretical uncertainty is automatically evaluated by {\sc HELAC-Onia}, as well as the factorisation- and renormalisation-scale uncertainties, evaluated from an independent variation in the interval\newline$\mu_F, \mu_R \in [1/2:2]\,\mu_0$, with $\mu_0 = m_T$. The mass uncertainty is obtained by varying $m_c$ in the range $m_c = 1.5 \pm 0.1$ GeV for all but the LO QED channel. For this latter channel, the invariant mass of the photon, $2 m_c$, should coincide with $M_{J/\psi}$ and we chose $m_c = 1.55$ GeV. We have also considered $\langle \mathcal{O}_{J/\psi}^{^{3\!}S_{1}^{[1]}} \rangle = 1.45$ GeV$^3$.

\begin{figure}[htp!]
\centering
\includegraphics[width=0.6\textwidth]{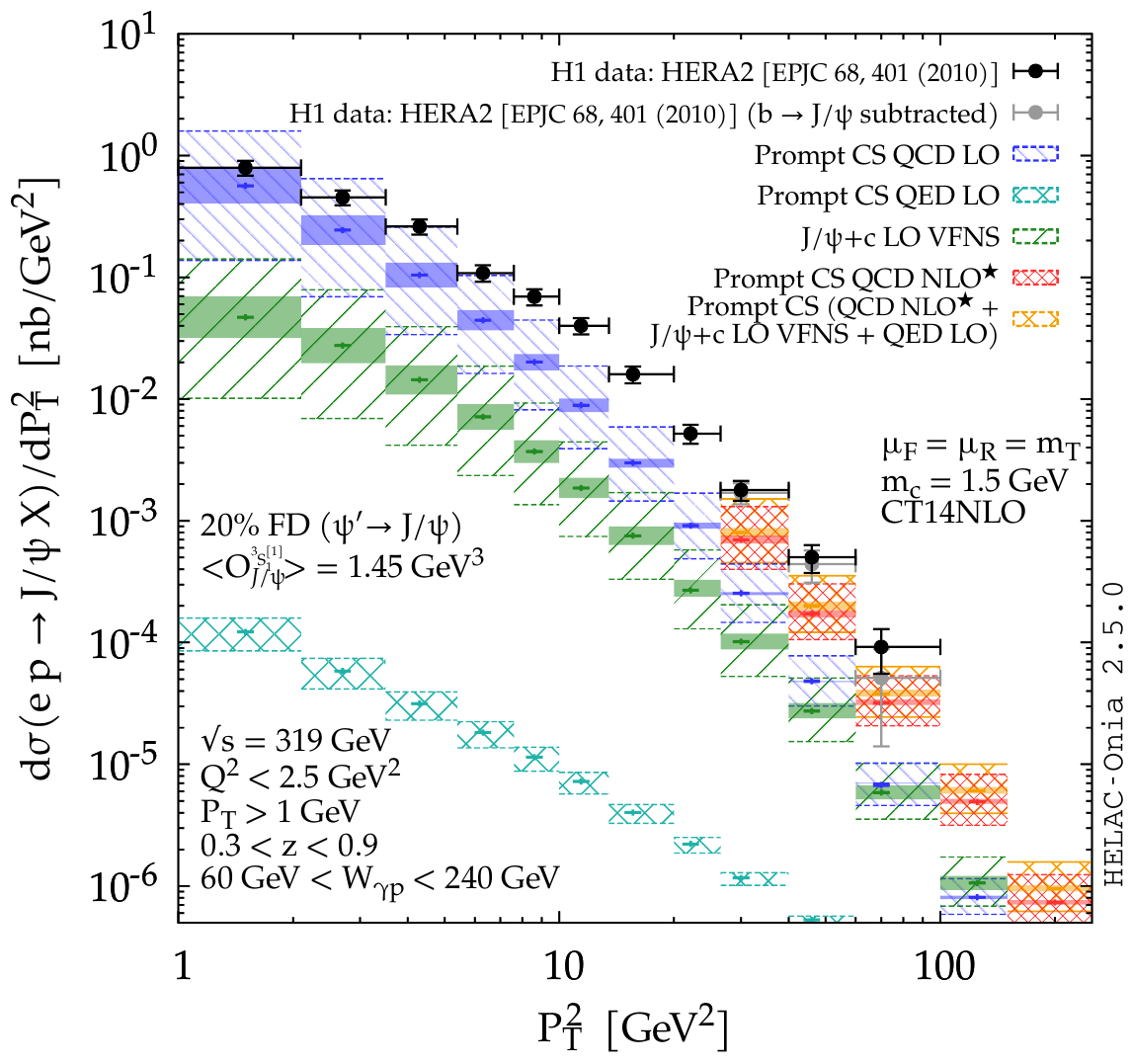}\vspace*{-.25cm}
\caption{Comparison between H1 data~\cite{Aaron:2010gz} (black: inclusive yield; grey: estimated prompt yield) and various CS contributions: LO (blue), \NLOstar (red), LO VNFS  $J/\psi+c$ (green), LO QED (light blue) and their combination (orange). The solid bands indicate the mass uncertainty while the patterns display the scale uncertainty. Figure taken from Ref.~\cite{Flore:2020jau}.
}
\label{fig:H1-NLOstar}
\end{figure}

We first start by revisiting HERA data. Our results are computed at H1 Run 2 kinematics~\cite{Aaron:2010gz}: $\sqrt{s_{ep}} = 319$ GeV, $Q^2 < 2.5$ GeV$^2$, $P_T > 1\,\text{GeV}$, $0.3 < z < 0.9$, $60\,\text{GeV} < W_{\gamma p} < 240\,\text{GeV}$. From Fig.~\ref{fig:H1-NLOstar}, we note the following: 1) the LO QCD contribution well describes the bulk of the data at low $P_T$; 2) the LO QED contribution is smaller in size, but its spectrum is harder with respect to the LO QCD one; 3) the LO VFNS $\jpsi+$charm contribution is not negligible and matters at large $P_T$; 4) the QCD \NLOstar is close to the data points, and the overall sum nearly agrees with them. Such an agreement is enhanced when subtracting the expected $b \to \jpsi$ feed-down from the data\footnote{See Appendix A of Ref.~\cite{Flore:2020jau} for a detailed discussion about the feed-downs from $b$ quarks.}. This being said, we argue that the CSM up to order $\alpha \alpha_s^3$ is able to reproduce HERA data. For this reason, we restrict our EIC predictions to the CSM.

Moving now to the EIC, we remark that this future collider will be able to run at different $\sqrt{s_{ep}}$ with very large luminosities. We consider two energy configurations: $E_e = 5\,(18)$ GeV and $E_p = 100\,(275)$ GeV, resulting in $\sqrt{s_{ep}} = 45\,(140)$ GeV. Also here, we apply some kinematical cuts: $P_T > 1$ GeV, $0.05 < z < 0.9$, $Q^2 < 1$ GeV$^2$; we also consider two different $W_{\gamma p}$ regions: $[10:40]$ GeV and $[20:80]$ GeV for $\sqrt{s_{ep}} = 45$ and $140$ GeV respectively.

\begin{figure}[htbp!]
\centering
\includegraphics[width=0.47\textwidth]{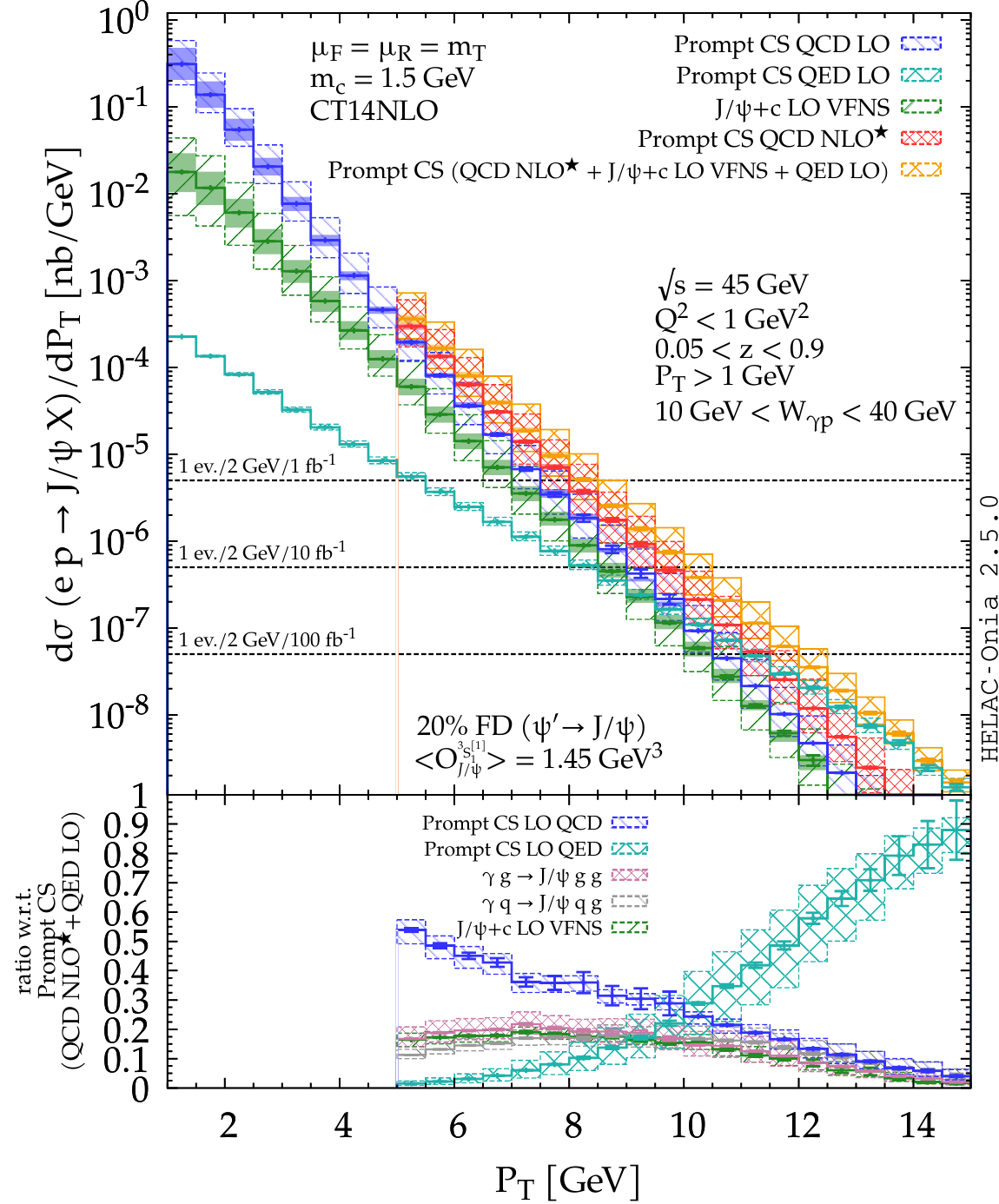}\label{fig:EIC-45GeV-NLOstar}
\includegraphics[width=0.475\textwidth]{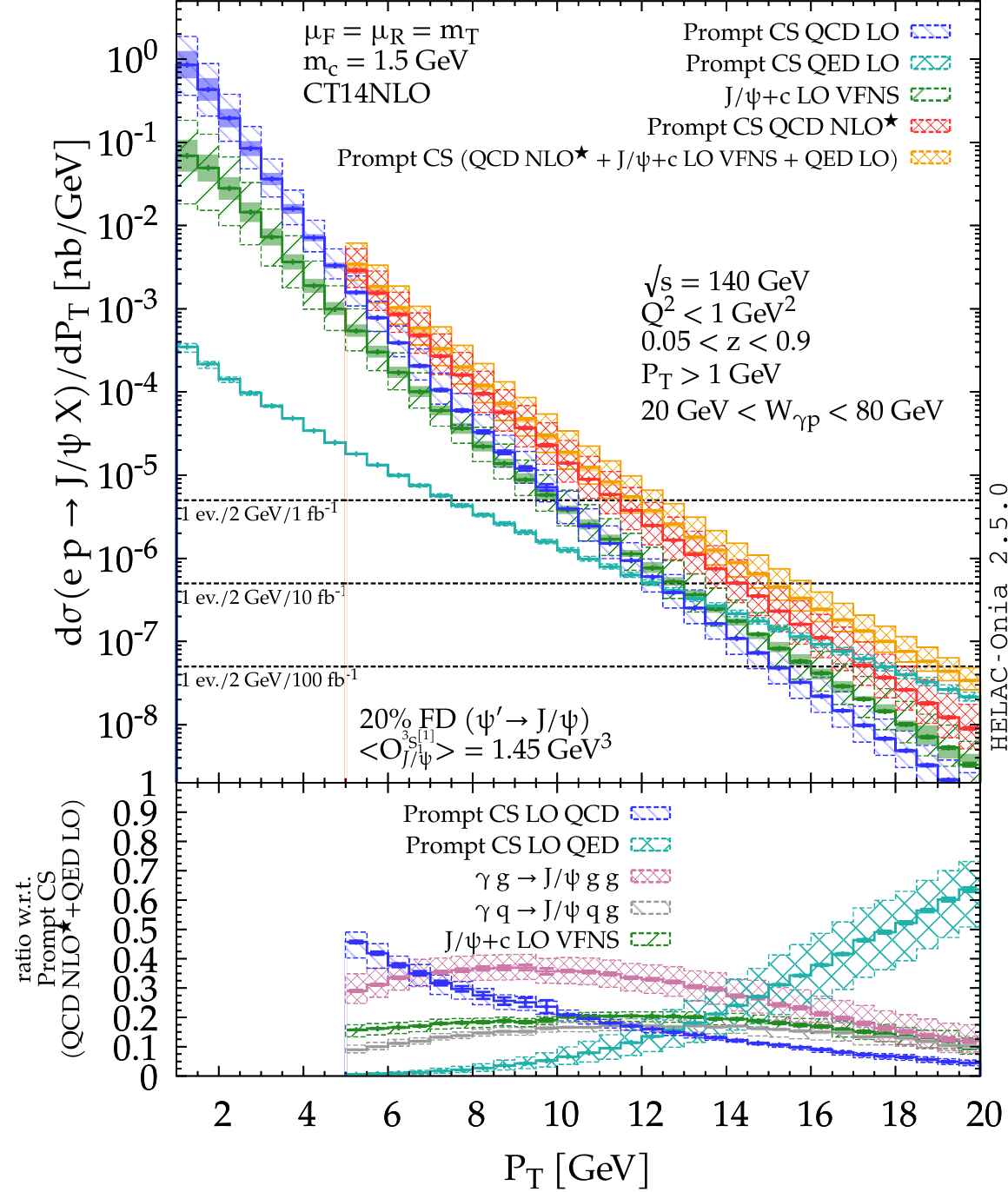}\label{fig:EIC-140GeV-NLOstar}
\caption{Predictions for the future EIC at $\sqrt{s_{ep}} = 45$ GeV (left) and $\sqrt{s_{ep}} = 140$ GeV (right). The calculation is performed adopting the same $\mu_F$, $\mu_R$ and PDFs as Fig.~\ref{fig:H1-NLOstar} with the same meaning for bands. Figure taken from Ref.~\cite{Flore:2020jau}.}
\label{fig:EIC-CT14NLO-NLOstar}
\end{figure}

By looking at Fig.~\ref{fig:EIC-CT14NLO-NLOstar}, we can note some differences between the two kinematical configurations. At $\sqrt{s_{ep}} = 45$ GeV (Fig.~\ref{fig:EIC-CT14NLO-NLOstar}, left panel), as $P_T$ increases, one enters the valence region, and the QED contribution becomes the dominant one at the largest measurable $P_T \simeq 11$ GeV at $\mathcal{L} = 100$ fb$^{-1}$. Moreover, $\gamma + q$ fusion contributes more than $30\%$ for $P_T > 8$ GeV, and the $\jpsi\,+$ unidentified charm contribution is comparable to the $\gamma + g(q)$ fusion subprocesses. Hence, both of the so far overlooked contributions will be relevant at the EIC. Looking then at $\sqrt{s_{ep}} = 140$ GeV (right panel in Fig.~\ref{fig:EIC-CT14NLO-NLOstar}), one can note that the yield is measurable up to $P_T \sim 18$ GeV. Also at this energy, the QED contribution is the leading one at the largest reachable $P_T$, while $\gamma + g$ fusion results the dominant contribution up to $P_T \sim 15$ GeV. More generally, it turns out that the production of $\jpsi+2$ hard partons (\ie $\,\jpsi+\{gg, qg, c\bar{c}\}$) is dominant for $P_T \sim 8 - 15$ GeV. This could lead to the observation of $\jpsi + 2$ jets with moderate $P_T$, with the leading jet$_1$ recoiling on the $\jpsi + {\rm jet}_2$ pair.

\subsection{$\jpsi\,+$ charm production}

Another interesting aspect that could be studied at the future EIC is the associated production of a $\jpsi$ and a charmed particle. In addition, one may also check to which extent the $\jpsi + c$ yield could help detecting a valence-like, non-perturbative charm content in the proton, to which we refer to as Intrinsic Charm (IC)~\cite{Brodsky:1980pb}. By considering the same LO VFNS computation used so far, but now including a $10\%$ charm detection efficiency $\varepsilon_c$ as follows:
\begin{equation}\label{eq:VFNS_with_efficiency}
 d\sigma^{\rm VFNS} = d\sigma^{\rm 3FS}\left[1-\left(1-\varepsilon_c\right)^2\right] + \left(d\sigma^{\rm 4FS} - d\sigma^{\rm CT}\right)\varepsilon_c
\end{equation}
(where $d\sigma^{\rm CT}$ is a proper counterterm to properly merge 3FS and 4FS contributions), we studied the $\jpsi + c$ yield at the two different EIC energy configurations. To do so, we employed the CT14nnloIC PDFs set~\cite{Hou:2017khm}, that encodes some eigensets with different IC effects: a ``sea-like'' one (in green) and a ``valence-like'' one, also called ``BHPS'' (in red). The central eigenset, to which we refer to as ``no IC'', is depicted in blue. Note that these IC effects are only affecting the 4FS contribution in Eq.~(\ref{eq:VFNS_with_efficiency}). 

\begin{figure}[t]
\centering
\includegraphics[width=0.475\textwidth]{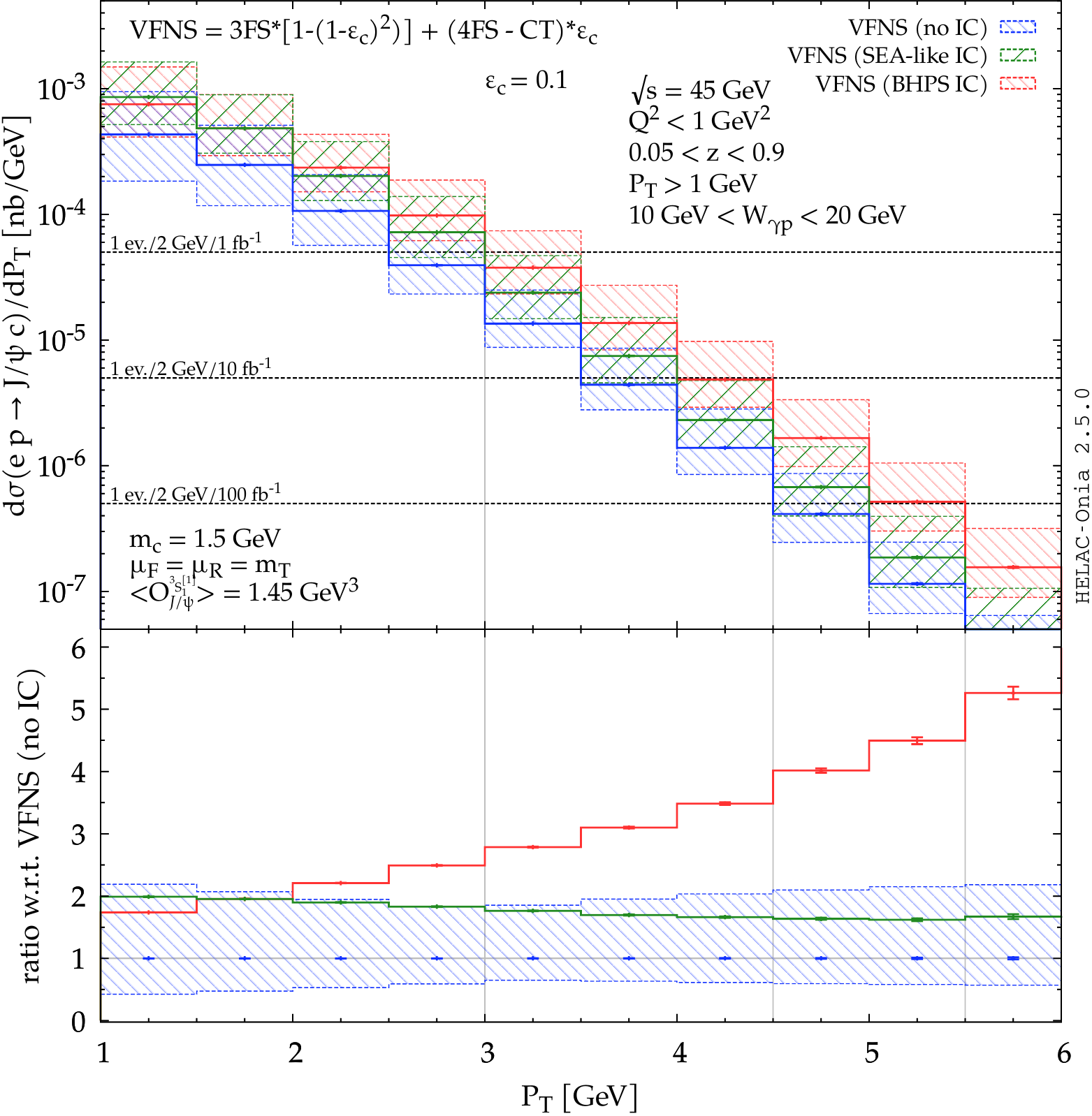}
\includegraphics[width=0.475\textwidth]{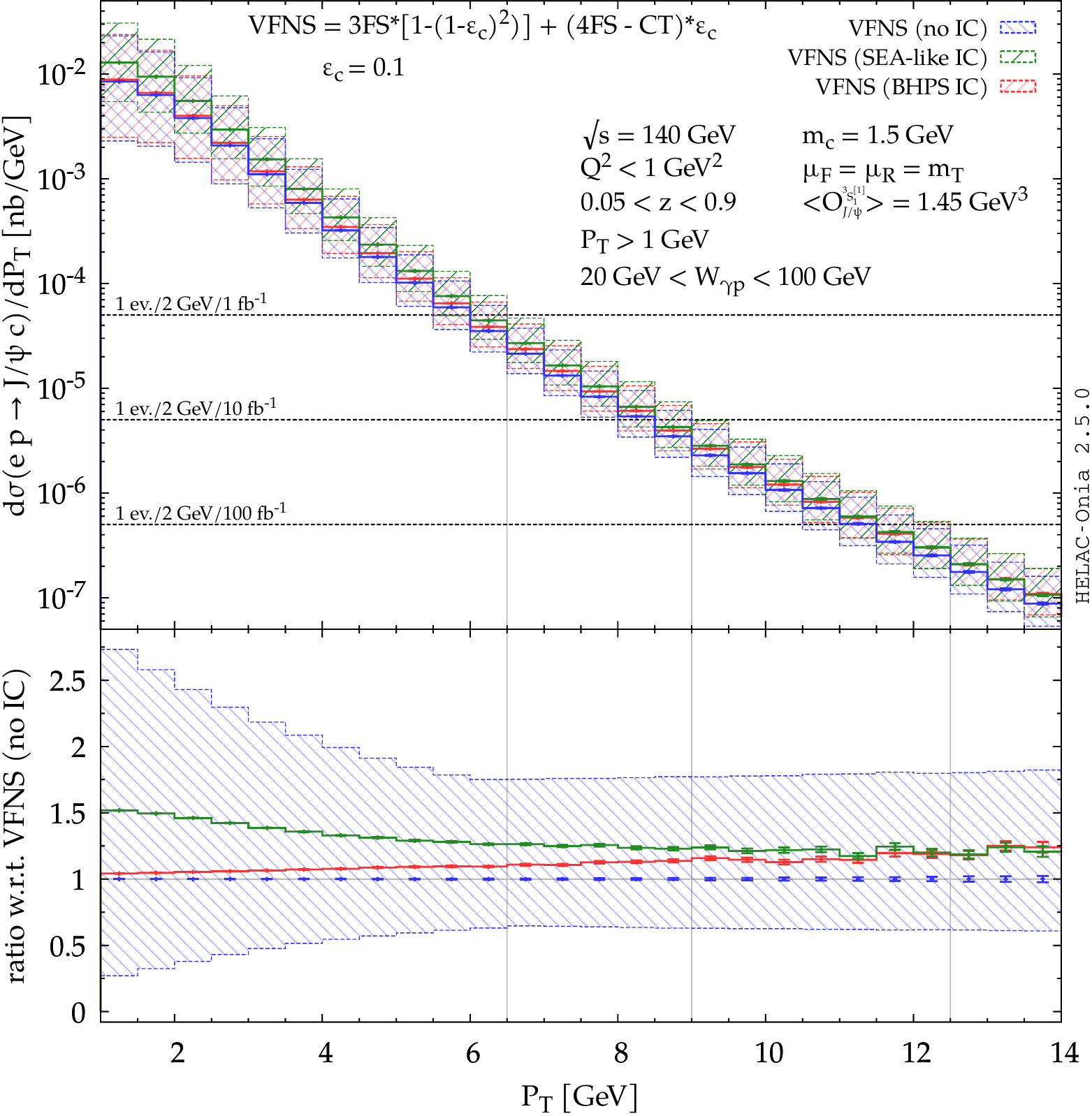}
\caption{Predictions for $\jpsi +$charm production at the future EIC at $\sqrt{s_{ep}} = 45$ GeV (left) and $\sqrt{s_{ep}} = 140$ GeV (right) computed with charm PDFs with ``no IC'' (blue), sea-like (green) and ``BHPS'' valence-like (red) ICs. The lower panel shows the ratio to the ``no IC'' curves and its relative uncertainty. The kinematical cuts are the same as in Fig.~\ref{fig:EIC-CT14NLO-NLOstar}. A $10\%$ charm detection efficiency is considered when drawing the horizontal observability lines. Figure taken from Ref.~\cite{Flore:2020jau}.}
\label{fig:EIC-VFNS-IC}
\end{figure}

By looking at Fig.~\ref{fig:EIC-VFNS-IC}, left panel, we see that at $\sqrt{s_{ep}} = 45$ GeV the $\jpsi + c$ yield is limited to low $P_T$ even with the largest integrated luminosity. Nonetheless, the yield is clearly observable if $\varepsilon_c = 0.1$ with $\mathcal{O} (500, 50, 5)$ events for $\mathcal{L} = (100,10,1)$ fb$^{-1}$. On the other hand, at $\sqrt{s_{ep}} = 140$ GeV (Fig.~\ref{fig:EIC-VFNS-IC}, right panel), the $P_T$ range is up to $10$ GeV, and we expect $\mathcal{O}(10^3)$ events at $\mathcal{L} = 100$ fb$^{-1}$. Such events could be observed by measuring a charmed jet.
Finally, note that as the valence region at high $x$ is not probed, no clear IC effect is visible at $\sqrt{s_{ep}} = 140$ GeV, while at $\sqrt{s_{ep}} = 45$ GeV, we observe a measurable effect, where the BHPS valence-like peak is visible.

\section{Conclusion}

We have analysed inclusive, large-$P_T$ $\jpsi$ photoproduction at $ep$ colliders. By including new, so far overlooked partonic subprocesses and feed-down contributions, we have shown that the CSM at $\mathcal{O}(\alpha \alpha_s^3)$ is able to describe the latest HERA data measured by the H1 Collaboration. We then provided predictions for the future EIC, where the new contributions, the $\mathcal{O}(\alpha^3)$ LO QED and the $\jpsi+$charm, can have a non-negligible role at this future collider. Furthermore, we showed that EIC can represent an ideal playground to constrain the proton charm content via the associated production of a $\jpsi$ and an identified charmed particle.
Such studies can be surely extended to other planned future colliders, such as the LHeC and FCC-eh~\cite{AbelleiraFernandez:2012cc, Agostini:2020fmq}.

\section*{Acknowledgements}
We thank Y.~Feng, M.A.~Ozcelik, J.W.~Qiu, I. Schienbein, H. Spiesberger for useful discussions. 

\paragraph{Funding information}

This project has received funding from the European Union's Horizon 2020 research and innovation programme under grant agreement No.~824093 in order
to contribute to the EU Virtual Access {\sc NLOAccess}.
This work  was also partly supported by the French CNRS via the IN2P3 project GLUE@NLO, via the Franco-Chinese LIA FCPPL (Quarkonium4AFTER), by the Paris-Saclay U. via the P2I Department and by the P2IO Labex via the Gluodynamics project, the French ANR under the grant ANR-20-CE31-0015 (PrecisOnium), and the CNRS IEA (GlueGraph). L.Y.~is supported by the EU Erasmus+ Paris-Saclay U.-Ukraine program.




\nolinenumbers

\end{document}